\documentclass[aps,prmaterials,twocolumn,longbibliography,superscriptaddress]{revtex4-1}

\usepackage{physics}
\usepackage{lmodern}
\usepackage{graphicx}
\usepackage{amsmath}
\usepackage{amsfonts}
\usepackage{amssymb}
\usepackage{bm}
\usepackage{xcolor}
\usepackage{hyperref}
\usepackage{ulem}
\usepackage{url}

\def\k{{ {\bm k} }}

\def\0{{ {\bm 0} }}

\def\ve{{\varepsilon}}

\def\rR{{ {\rm R} }}
\def\rA{{ {\rm A} }}

\def\cV{{ {\cal V} }}

\allowdisplaybreaks[4]

\begin{document}

\title{Orbital-related gyrotropic responses in Cu$_2$WSe$_4$ and chirality indicator}
\author{Kazuki Nakazawa} 
\affiliation{RIKEN Center for Quantum Computing, Wako 351-0198, Japan}
\author{Terufumi Yamaguchi} 
\affiliation{Department of Physics, Kobe University, Kobe 657-8501, Japan}
\affiliation{RIKEN Center for Emergent Matter Science, Wako 351-0198, Japan}
\author{Ai Yamakage}
\affiliation{Department of Physics, Nagoya University, Nagoya 464-8602, Japan}
\date{\today}

\begin{abstract}
In recent years, counterparts of phenomena studied in spintronics have been actively explored in the orbital sector. The relationship between orbital degrees of freedom and crystal chirality has also been intensively investigated, although the distinction from gyrotropic properties has not been fully clarified. In this work, we investigate spin and orbital Edelstein effects as well as the nonlinear responses in the ternary transition-metal chalcogenide Cu$_2$WSe$_4$, which has a gyrotropic but achiral crystal structure. We find that in the Edelstein effect, magnetization is dominated by the orbital contribution rather than the spin contribution. On the other hand, both the nonlinear chiral thermoelectric (NCTE) Hall effect—a response to the cross product of the electric field and the temperature gradient—and the nonlinear Hall effect—conventional second-order response to the electric field—are found to be dominated by the Berry curvature dipole. We further find that spin-orbit coupling plays only a minor role in these effects, whereas the orbital degrees of freedom are essential. Finally, we demonstrate that the orbital magnetic-moment contributions to both the Edelstein effect and the NCTE Hall effect are closely linked to chirality, and we discuss the possibility of using them as a chirality indicator.
\end{abstract}

\maketitle

\section{Introduction}
Electron spin underlies magnetism, and its manipulation is central to spintronics~\cite{ZFS2004,Hirohata2020} and quantum information devices~\cite{LD1998,HKPTV2007,KL2013}. At the same time, spin angular momentum can be converted into orbital angular momentum through spin–orbit coupling (SOC)~\cite{DP1971PLA,DP1971JETP,Wolf2001,MNZ2003,GG2014}. Consequently, concepts long developed for spin have been extended to the orbital sector~\cite{Go2021,Jo2024,Ando2025}: orbital‑related quantities such as orbital magnetization~\cite{CN1996,SN1999,XSN2005,TCVR2005} and orbital currents~\cite{BHS2005,KTHYI2008,Tanaka2008,Choi2023} are now active subjects of study.

A paradigmatic example in which orbital magnetization plays a central role is current‑induced orbital magnetization—the orbital Edelstein effect~\cite{Park2011,Park2012,Yoda2015,ZNS2016,Yoda2018}. In noncentrosymmetric conductors, an electric field drives a nonequilibrium magnetization, i.e., a magnetoelectric cross‑correlation that converts an electrical input into a magnetic output~\cite{ALG1991,Edelstein1990,Sanchez2013,GG2014,Lesne2016,FSKI2017,Johansson2024}. While early work focused on the spin‑based Edelstein effect, recent theory and experiments have established a purely orbital counterpart that can even dominate in appropriate materials~\cite{Park2011,Park2012,Yoda2015,ZNS2016,Yoda2018,ElHamdi2023,Krishnia2024,Gobel2025cp,GSM2025}. Importantly, the orbital Edelstein effect does not require SOC, suggesting routes to efficient charge–magnetization interconversion in light‑element platforms~\cite{Yoda2015,Yoda2018} and providing possible explanation of the chirality-induced spin/orbital selectivity~\cite{Gobel2025cp,GSM2025}.

Beyond linear response, second‑order responses to the electric field and/or thermal gradient have emerged as a powerful probe of band geometry and exotic magnetic structure~\cite{Sipe2000,GYN2014,SF2015,TSM2018,Hidaka2018,Nakai2019,Toshio2020,PMOM,JL,MP,DLX2021,Li2021,NKM2022,Legg2022,OK2022,DLBLK,GCC2023,AXC2023,DLACA2023,YNY2023,NYY2024,YNY2024,Arisawa2024,NYY2025,NLKL2025,Arisawa2025,Nomoto2025}. Moreover, when an electric field and a temperature gradient are applied perpendicular to one another, a transverse current can appear along the direction of their cross product—the nonlinear chiral thermoelectric (NCTE) Hall effect~\cite{Hidaka2018,Nakai2019,Toshio2020,YNY2023,NYY2024,NYY2025,YNY2024,Nomoto2025}. In nonlinear transverse responses such as the nonlinear Hall effect, a Berry‑curvature dipole is often essential; for the NCTE Hall effect, there is, in addition, a contribution from the orbital magnetic moment, whose connection to the orbital Edelstein effect has been discussed~\cite{YNY2023}. 

These higher-order responses, as well as the magnetoelectric cross-correlation, require specific crystal symmetries to occur. Indeed, both the Edelstein and NCTE Hall effects are permitted precisely in gyrotropic crystals—systems in which a polar vector $\bm P$ and an axial vector $\bm A$ are linearly related by a second‑rank gyrotropic tensor ${\cal G}$, $P_i={\cal G}_{ij}A_j$, and for which the point‑group symmetry allows ${\cal G}\neq 0_{3 \times 3}$  ($3\times 3$ zero tensor)~\cite{GG2014,FSKI2017,HL2020,SW2025}. Within this class, crystals that additionally lack any rotoreflection axis are chiral. Although numerous studies have hinted at links between orbital magnetic moments and (crystal) chirality, decisive evidence remains scarce, and recent efforts have turned to quantifying chirality itself~\cite{MISH2025,OK2025}. Clarifying how transport coefficients reflect gyrotropy versus chirality is therefore a key step toward understanding chiral materials.

\begin{figure*}
\hspace*{0mm}
\includegraphics[width=175mm]{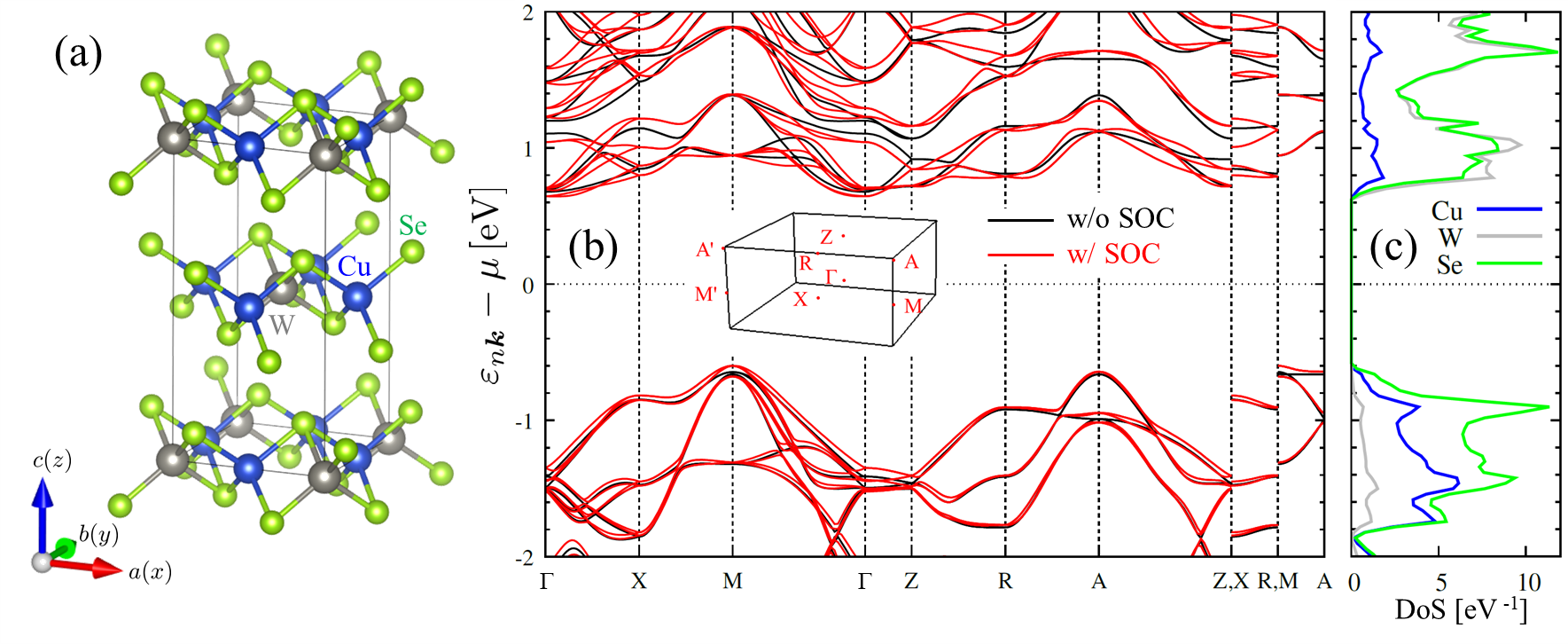}
\vspace*{-5mm}
\caption{
(a) Crystal structure of Cu$_2$WSe$_4$. The cuboid indicates the conventional unit cell, which contains two primitive cells. (b) Band structure with SOC (red) and without SOC (black). The inset shows the Brillouin zone and high‑symmetry points. (c) Projected partial density of states (DOS) from the relativistic calculation for each atomic species.}
\label{fig:1}
\end{figure*}

Here, we propose a concrete platform to disentangle these notions. The ternary transition‑metal chalcogenides ${\rm Cu}_2MX_4$ ($M=$ Mo, W; $X=$ S, Se, Te) crystallize in the noncentrosymmetric space group $I\bar{4}2m$ (No.~121) [or $P\bar{4}2m$ (No.~111), not considered in this paper] and comprise stacked layers of edge‑sharing Cu$X_4$ and $MX_4$ tetrahedra; each $X$ atom is coordinated by two Cu and one $M$ atoms [Fig.~\ref{fig:1}(a)]~\cite{CHE2005,GS2014,Chen2015,YSO2018,Nadeem2022,Nbeg2023}. The point group is $D_{2d}$, rendering the structure gyrotropic yet achiral. We focus on Cu$_2$WSe$_4$, a nonmagnetic wide‑gap semiconductor. Prior work has reported its optical~\cite{Kocyigit2019,Nadeem2022,Nbeg2023} and elastic~\cite{Nbeg2023} properties, as well as relatively high thermoelectric power factors and figures of merit ($ZT$)~\cite{Nadeem2022}. However, systematic studies of phenomena that rely on its noncentrosymmetry—such as the Edelstein effect and nonlinear transport—are still lacking, presenting an opportunity for both fundamental and applied exploration. Moreover, by contrasting this achiral, gyrotropic platform with genuinely chiral counterparts, one can isolate which aspects of the observed responses stem from gyrotropy alone.

In this paper, we examine the electronic structure and orbital properties of the achiral gyrotropic semiconductor Cu$_2$WSe$_4$, and then discuss the general feature on the relation between chirality and orbital magnetic moment. We find that the orbital Edelstein effect dominates over the spin contribution. For the nonlinear Hall and NCTE Hall effects, we find that contributions from the Berry curvature are dominant, while those from the orbital magnetic moment are subdominant. We further discuss common properties of the response coefficients that characterize the Edelstein and NCTE Hall effects, and we clarify distinctions between chiral and gyrotropic systems by introducing \lq\lq chirality indicator." 

The remainder of this paper is organized as follows. In Sec.~\ref{sec:band}, we investigate the electronic structure of Cu$_2$WSe$_4$ using first-principles calculations and discuss its orbital properties. In Sec.~\ref{sec:EN}, employing a Wannier-based model, we calculate the spin/orbital Edelstein effects, the nonlinear Hall effect, and the NCTE charge and thermal Hall effect, and analyze their characteristics. In Sec.~\ref{sec:GC}, we discuss quantities that characterize chirality and gyrotropy using transport coefficients (tensors) that encompass both the Edelstein and NCTE Hall effects. Sec.~\ref{sec:conc} concludes this paper. 

\section{Band structure}
\label{sec:band}

We use OpenMX code~\cite{OpenMX,Ozaki2003} to obtain the band structure based on the density functional theory (DFT). The wave functions are expanded using linear combinations of pseudoatomic orbitals. Generalized gradient approximation (GGA) proposed by Perdew-Burke-Ernzerhof~\cite{PBE1996} is used for the exchange-correlation functional, and norm-conserving and total angular momentum-dependent pseudopotentials are chosen. We perform a fully relativistic calculation when including SOC. The basis set for pseudoatomic orbitals is employed as Cu6.0H-s3p2d1, Se7.0-s3p2d2, and W7.0-s3p2d2f1. A conventional unit cell (Cu$_4$W$_2$Se$_8$) is employed to adjust the $c$ axis to the $S_4$ rotoreflection axis. We use the lattice constants of $a=b=5.560$~\AA ~and $c=11.214$~\AA. We set the cutoff energy, which specifies the fast Fourier transform grid, to 1200 Ry and sampled the Brillouin zone with $16^3$ $k$-point mesh. The self-consistent field calculation converged to the paramagnetic state, which agrees with the previous report~\cite{Nadeem2022}. 

Figure~\ref{fig:1}(b, c) shows the band structure and the partial density of states. DFT-GGA calculation yields an indirect band gap of $\sim 1.2$ eV ($\Gamma \to {\rm M}$), consistent with a previous study~\cite{Nadeem2022}. This slightly underestimates the experimental value obtained by the optical absorption measurements ($1.49$--$1.64$ eV)~\cite{YSO2018,Kocyigit2019} because electron correlations are neglected~\cite{Nadeem2022}, which is unnecessary for our purpose. We can see that the valence bands are dominated primarily by Se-$p$ and Cu-$d$, while the conduction bands are dominated by W-$d$ and Se-$p$ states. Due to the strong SOC of W, larger spin-orbit splitting in the conduction band is observed compared to that in the valence band.

\section{Edelstein effect and nonlinear transport properties}
\label{sec:EN}

\begin{figure}
\hspace*{0mm}
\includegraphics[width=75mm]{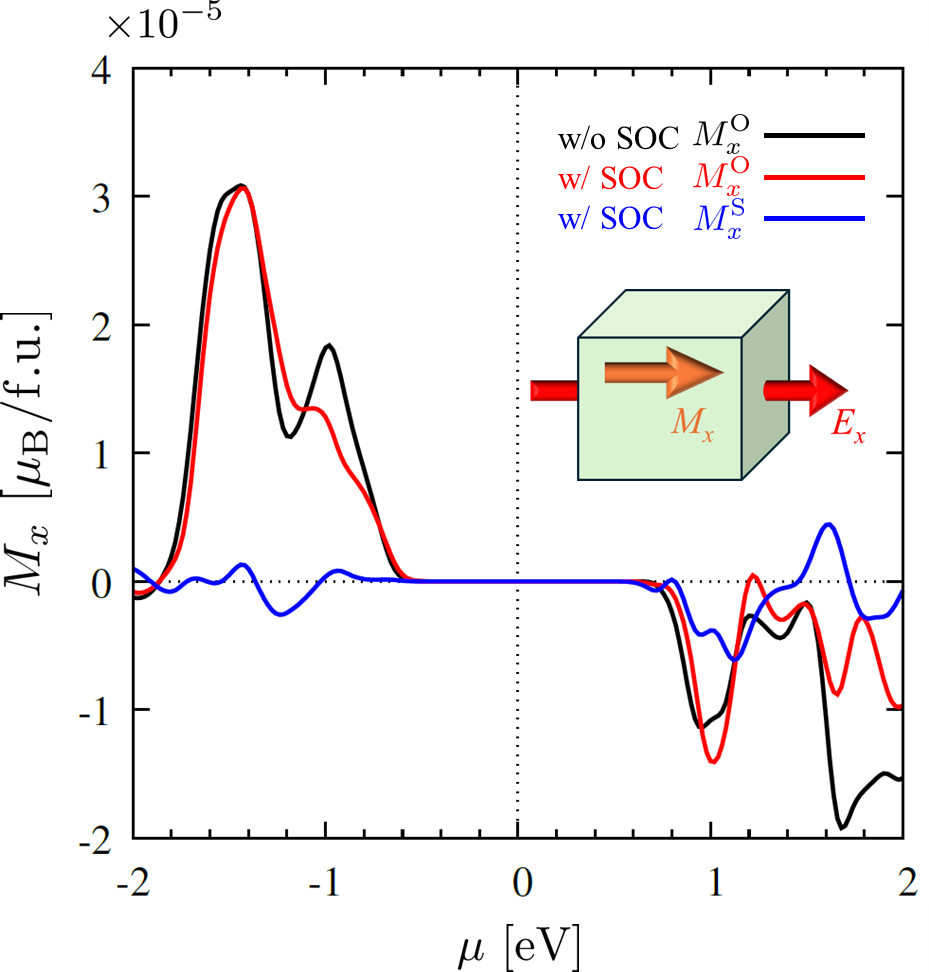}
\caption{Orbital Edelstein magnetization without (black line) and with (red line) SOC and the spin Edelstein magnetization (blue line).  
}
\label{fig:2}
\end{figure}

For the calculation of each physical quantity, we use OpenMX~\cite{OpenMX,WOT2009} to construct maximally localized Wannier functions. The DFT bands within the energy range from $-7$~eV to +2.45~eV are projected to 108-orbitals, including spin degrees of freedom, and consist of Cu-$d$, W-$d$, and Se-$p$ orbitals. The Wannier model almost perfectly reproduces the original DFT band. 

We calculated the spin/orbital Edelstein magnetization, NCTE charge and thermal Hall currents, and nonlinear Hall conductivity using the obtained Wannier model. We employ sufficiently dense $k$-meshes of $180^3$-$400^3$ for the momentum integrals. 

\subsection{Spin and orbital Edelstein effects}
The Edelstein magnetization is given as $M_i = \alpha_{ij} E_j = M_i^{\rm S} + M_i^{\rm O}$, where~\cite{Yoda2015,ZNS2016,Yoda2018,Johansson2024} 
\begin{gather}
\alpha_{ij} = \frac{e\tau}{V} \sum_{n\k} \left( -\frac{\partial f}{\partial \ve} \right)_{\ve=\ve_{n\k}} \left( {\bm m}^{\rm S} + {\bm m}^{\rm O} \right)_{n\k}^i v_{j,n\k } ,
\\
{\bm m}_{n\k}^{\rm S} = -\frac{g\mu_{\rm B}}{2} \langle n \k | {\bm \sigma} | n \k \rangle, 
\\
{\bm m}_{n\k}^{\rm O} = - \frac{e}{2\hbar} {\rm Im} [ \langle \bm \nabla_\k n (\k) | \times \{ \hat{H}_\k - \ve_{n \k} \} | \bm \nabla_\k  n (\k) \rangle ] . 
\end{gather} 
Here we imply the dc electric field $\bm E$, an electron charge $e<0$, an electron lifetime $\tau$, a system volume $V$, the Fermi-Dirac distribution function $f \equiv f(\ve)$, the group velocity ${\bm v}_{n\k} = \frac{1}{\hbar} {\bm \nabla}_\k \ve_{n\k}$ with an eigenenergy $\ve_{n\k}$ and an eigenvector $| n (\k) \rangle$ of the Hamiltonian $\hat{H}_\k$, and spin (orbital) magnetic moment ${\bm m}_{n\k}^{\rm S(O)}$ with the spin $g$-factor $g = 2$, Bohr magneton $\mu_{\rm B}$, and Dirac constant $\hbar$. 

In Fig.~\ref{fig:2}, we present plots of the spin and orbital components of the Edelstein magnetization. An electric field is applied along the $x$ direction, and only the induced component $\alpha_{xx}$ is shown. This is because the point-group symmetry of this system is $D_{2d}$, and one can show that $\alpha_{xx} = -\alpha_{yy}$ and all other components vanish, meaning that there is only one independent component (see Appendix~\ref{sec:G}). The calculation assumes a dc electric field strength of $E_x = 10^4$~V/m, an electronic scattering rate of $\gamma = \hbar/(2\tau) = 30$ meV, and a temperature of $k_{\rm B} T = 30$~meV. We find that a measurable Edelstein magnetization is induced under a reasonable strength of the electric field in this system. We also find that the spin Edelstein magnetization contributes comparably to that of orbital one in the conduction band because of the strong SOC of tungsten, while the spin component is largely suppressed and the orbital component dominates in the valence band due to the weaker SOC of copper. 

To further examine the impact of SOC, we compared orbital Edelstein magnetization with and without SOC. In the conduction bands, it is seen that the orbital Edelstein magnetization is relatively suppressed due to the strong SOC of tungsten. Apart from this suppression, however, the overall behavior of the orbital Edelstein magnetization is well captured by a model without SOC. In particular, in the valence bands, the results with and without SOC agree well. These results indicate that the importance of orbital nature for the emergence of the Edelstein effect, and that the SOC is not an essence in this material. These results represent the potential of the present Cu$_2$WSe$_4$ for orbitronics devices. 

\begin{figure*}
\includegraphics[width=160mm]{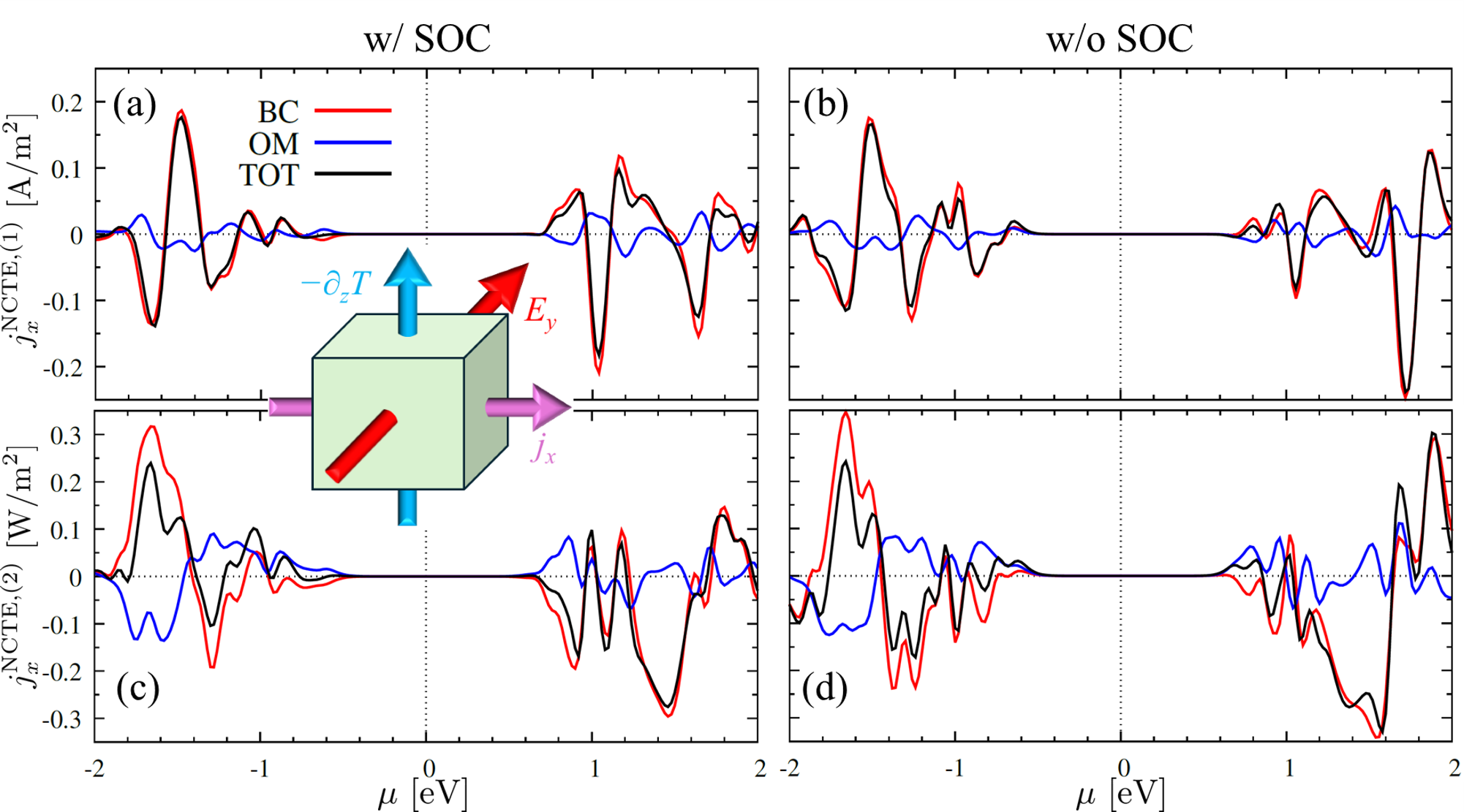}
\caption{[(a) and (b)] NCTE charge Hall current (a) with SOC and (b) without SOC. [(c) and (d)] NCTE thermal Hall current (c) with SOC and (d) without SOC. Berry curvature (BC) contribution (red lines), orbital magnetic moment (OM) contribution (blue lines), and total value (black lines) are plotted in each panel. 
}
\label{fig:3}
\end{figure*}

\subsection{NCTE charge and thermal Hall effects}
We introduce the expression of NCTE charge ($\ell=1$) and thermal ($\ell=2$) Hall current $j_i^{{\rm NCTE},(\ell)} = \chi_{ij}^{(\ell)} \{ {\bm E} \times (- {\bm \nabla} T/T ) \}_j$. In this form, the second-rank tensor $\chi_{ij}^{(\ell)}$ satisfies the same requirement to $\alpha_{ij}$, that is, $\chi_{xx}^{(\ell)} = -\chi_{yy}^{(\ell)}$ and otherwise zero. 
The microscopic calculation revealed that the NCTE charge and thermal Hall current are dominated by the following two terms~\cite{YNY2023,NYY2024,NYY2025};  
\begin{align} 
&\chi_{xx}^{(\ell)} \simeq \chi^{\mathrm{BC,}(\ell)}_{xx} + \chi^{\mathrm{OM,}(\ell)}_{xx} , \label{eq:NCTE_TOT} \\
&\chi^{\mathrm{BC,}(\ell)}_{xx} = \frac{e^{2} \tau}{\hbar} \frac{1}{V} \sum_{n,\bm{k}}
F_\ell(\ve_{n\k})
\left\{ 	\Omega'_x - \frac{1}{2} \left( 
	\Omega'_y + \Omega'_z \right) \right\} ,
\label{eq:NCTE_BC}
\\
&\chi^{\mathrm{OM,}(\ell)}_{xx} = -\frac{e\tau}{2\hbar} \frac{1}{V} \sum_{n,\bm{k}}
F_\ell (\ve_{n\k})
\bm{\nabla}_{\bm{k}} \cdot \bm{m}^{{\rm O},\perp}_{n\bm{k}} ,
\label{eq:NCTE_OM}
\end{align}
where $ F_\ell (\ve) = e^{1-\ell} (\ve - \mu)^\ell (-\frac{\partial f}{\partial \ve})$ with the temperature $T$, chemical potential $\mu$, $\Omega'_i \equiv v_{i,n{\bm k}} \Omega_{n\bm k}^i$, and ${\bm m}_{n\k}^{{\rm O},\perp} = \left( 0, m_{n\k}^{{\rm O},y}, m_{n\k}^{{\rm O},z} \right)$ is the orbital magnetic moment which only contain components perpendicular to the NCTE Hall current. The Berry curvature ${\bm \Omega}_{n\k}$ is calculated using
\begin{align}
{\bm \Omega}_{n\k} = - {\rm Im} [ \bm \nabla_\k \times \langle n (\k) | \bm \nabla_\k  n (\k) \rangle ] . 
\label{eq:OM}
\end{align} 
This expression is specific to the current along $x$; the $y$ component follows by cyclic permutation $x \to y \to z \to x$ and redefining ${\bm m}_{n\k}^{{\rm O},\perp} = \left( m_{n\k}^{{\rm O},x}, 0, m_{n\k}^{{\rm O},z} \right)$. 

In Fig.~\ref{fig:3}, we show the chemical potential dependence of the NCTE charge and thermal Hall effects. We set the dc electric field and the temperature gradient as $E_y = 1000$~V/m and $\partial_z T/T = 100$~m$^{-1}$; the damping rate and temperature are set as $\gamma = \hbar/(2\tau) = 30$ meV and $k_{\rm B} T = 30$~meV, respectively. Overall, it can be seen that the contribution from the Berry curvature dipole dominates over the orbital magnetic moment terms. This is partly attributed to the momentum-space structure of the Berry curvature and the orbital magnetic moment. By considering the $S_4$ rotoreflection [$S_4(X,Y,Z)^{\rm T} = (-Y, X, -Z)^{\rm T}$], one can show that $\sum_\k F_\ell \Omega_x' = -\sum_\k F_\ell \Omega_y'$ and $\sum_\k F_\ell \Omega_z' = 0$ for the Berry curvature, and the same applies to the orbital magnetic moment. Considering together with Eqs.~\eqref{eq:NCTE_BC} and \eqref{eq:NCTE_OM}, $\Omega_x'$ and $\Omega_y'$ contributes additively in the Berry curvature dipole term: $\sum_\k F_\ell ( \Omega_x' - \Omega_y'/2 ) =  (3/2) \sum_\k F_\ell \Omega_x'$, while only $y$ component contributes in the orbital magnetic moment term $(1/2) \sum_\k F_\ell \bm{\nabla}_{\bm{k}} \cdot \bm{m}^{{\rm O},\perp}_{n\bm{k}} = (1/2) \sum_\k F_\ell \partial_{k_y} m_{n\k}^{{\rm O},y}$. The consequences of $\sum_\k F_\ell \Omega_z' = \sum_\k F_\ell \partial_{k_z} m_z = 0$ are the absence of the monopole structures of the Berry curvature and the orbital magnetic moment, which is related to the achiral nature of the system. This point will be discussed in more detail in a later section, and it is also important for the nonlinear Hall effect discussed below.

Next, we discuss the effect of SOC. When SOC is taken into account, the electronic structure changes due to the spin--orbit splitting [see Fig.~\ref{fig:1}(b)], especially in the conduction bands, and accordingly, the overall behavior also changes. On the other hand, the typical magnitude of the NCTE charge and thermal currents—for example, their peak values—does not change so much. Based on these results, we again see that the effect of SOC is limited in this material, and that the Berry curvature contribution is mainly governed by orbital-crossing effects.

\subsection{Nonlinear Hall effect}

\begin{figure*}
\includegraphics[width=160mm]{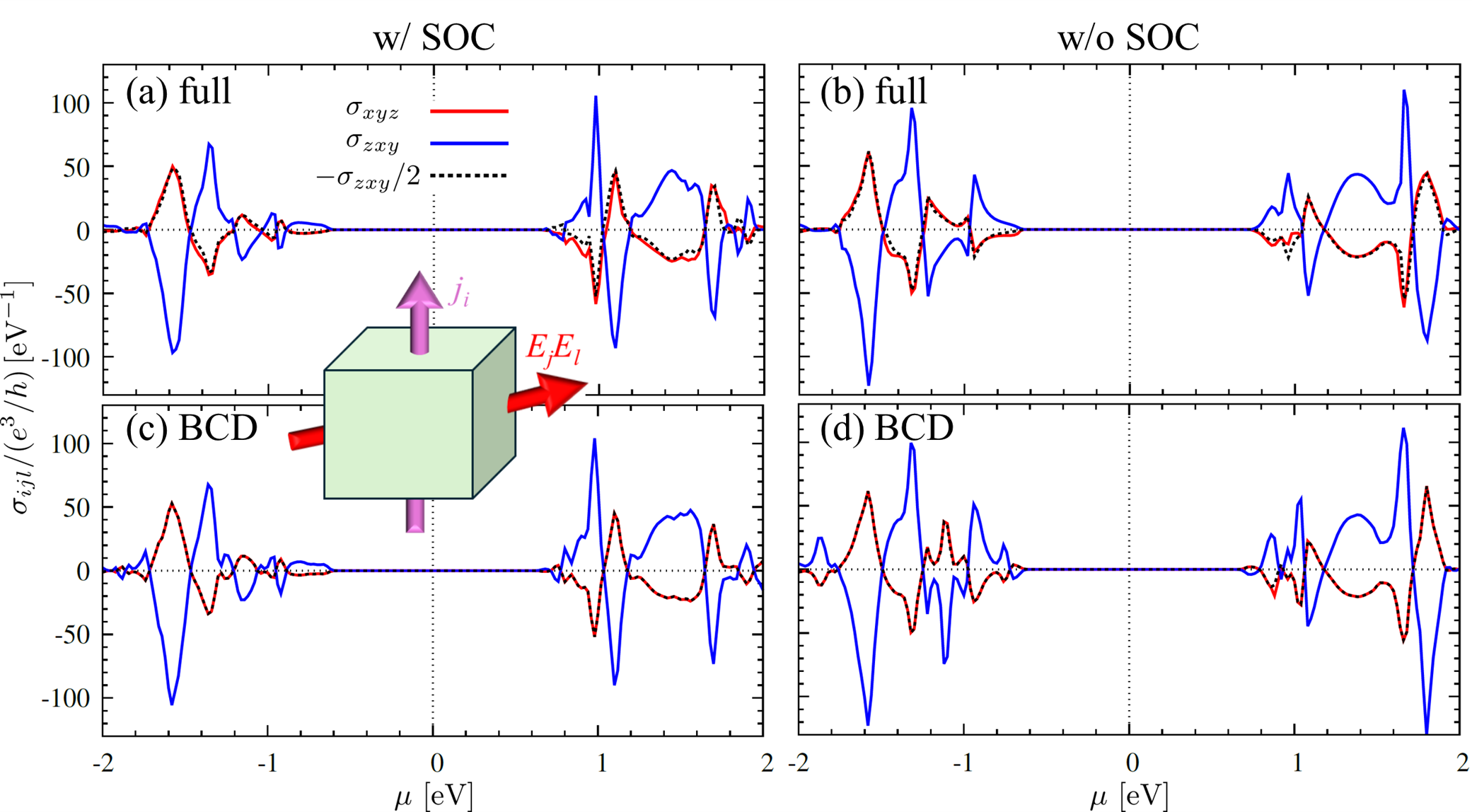}
\caption{Normalized nonlinear Hall conductivity $\sigma_{ijk}/(e^3/h)$ ($h$: Planck constant) calculated using (a,b) Eq.~\eqref{eq:NLC} (full) and (c,d) Eq.~\eqref{eq:nlbcd} (BCD), in the (a,c) presence and (b,d) absence of SOC. $xyz$ (red solid lines) and $zxy$ (blue solid lines) are plotted, and $-\sigma_{zxy}/2$ (black dotted lines) is also shown to compare with $xyz$ component. 
}
\label{fig:4}
\end{figure*}

The second-order current response to the dc electric field $j_i^{\rm NLC} = \sigma_{ijl} E_j E_l$ is also calculated. The second-order dc nonlinear conductivity $\sigma_{ijl}$ is given by~\cite{MP,MN,NYY2024,NLKL2025}
\begin{align}
&\sigma_{ijl} \simeq
\frac{e^3}{\hbar} \frac{1}{V} \int \frac{d\ve}{2\pi} \left(-\frac{\partial f}{\partial \ve}\right)  
\nonumber \\
&\quad \times {\rm Im}   \sum_\k 
{\rm tr} \Biggl\{ \hat{\cV}_i  
 \frac{\partial \hat{G}^\rR }{\partial \ve} \left( \hat{\cV}_j \hat{G}^\rR \hat{\cV}_l + \frac{1}{2} \hat{\cV}_{jl} \right) ( \hat{G}^\rR - \hat{G}^\rA ) \Biggr\}  
\nonumber \\
&\quad + (j \leftrightarrow l), 
\label{eq:NLC}
\end{align}
where $\hat{G}^\rR = (\ve - \hat{H}_\k - \hat{\Sigma}^\rR)^{-1} = (\hat{G}^\rA)^\dagger$ is retarded Green's function with self-energy $\hat{\Sigma}^{\rm R}$, $\hat{\cal V}_i = \partial_{k_i} \hat{H}_\k$, and $\hat{\cV}_{ij} = \partial_{k_i} \partial_{k_j} \hat{H}_\k$. The trace runs over all of the orbital/band indices. For simplicity, we here consider the constant pure imaginary self-energy $\hat{\Sigma}^{\rm R} = -i\hbar/(2\tau) = - 15i$~meV and evaluate at zero temperature. 

Symmetry constraints for $D_{2d}$ point group predict only two nonzero independent tensor components: $\sigma_{xyz} = \sigma_{yzx}$ and $\sigma_{zxy}$. We first numerically compute these components using Eq.~\eqref{eq:NLC} and obtain the result which approximately satisfies the relation $\sigma_{xyz} = \sigma_{yzx} \approx -\sigma_{zxy}/2$, as shown in Figs.~\ref{fig:4}(a,b). Given that the effect from Berry curvature was prominent also in the case of the NCTE charge and thermal Hall effects, we here suppose that the dominant contribution also comes from the Berry curvature. The Berry curvature dipole contribution is denoted as~\cite{SF2015,Nakai2019}:
\begin{equation}
\sigma_{ijl}^{\rm BCD} = \frac{e^2 \tau}{2\hbar} \frac{1}{V} \sum_{\bm k} F_0 (\varepsilon_{\bm k}) 
\left( \Omega_j' - \Omega_l' \right).
\label{eq:nlbcd}
\end{equation} 
Here, $i \neq j \neq l \neq i$ is assumed. Similarly, one can show $\sum_\k F_0 (\ve_{n\k}) \Omega_x' = -\sum_\k F_0 (\ve_{n\k}) \Omega_y'$ and $\sum_\k F_0 (\ve_{n\k}) \Omega_z' = 0$, leading to the relation $\sigma_{xyz} = \sigma_{yzx} = -\sigma_{zxy}/2$. We calculated the nonlinear Hall conductivities using Eq.~\eqref{eq:nlbcd} in the temperature of $k_{\rm B} T= 10~{\rm meV}$ and see the nice agreement with the full formula [see Figs.~\ref{fig:4}(c,d)]. 

Again, we discuss the effect of SOC by comparing Figs.~\ref{fig:4}(a,b) and (c,d). Similarly to Edelstein and NCTE Hall effect, we confirmed that the approximate magnitude of the nonlinear conductivity is not significantly affected by SOC.

\section{Gyrotropic and chiral properties}
\label{sec:GC}

\begin{table}[t]
\centering
\newlength{\height} 
\setlength{\height}{3mm}
\begin{center}
\fontsize{10pt}{10pt}\selectfont
\begin{tabular}{ccccccll}
\hline\hline
 $D_{2d}$ & $E$ & $2S_4$ & $C_2$ & $2C_2'$ & $2\sigma_d$ & Linear & Quadratic
\\
\hline
 $A_1$ & 1 &  1   &  1   &  1   &  1   &       & $XX+YY, \, ZZ$
\\
 $A_2$ & 1 &  1   &  1   & $-1$ & $-1$ & $R_Z$ & $XY-YX$
\\
 $B_1$ & 1 & $-1$ &  1   &  1   & $-1$ &       & $XX-YY$
\\
 $B_2$ & 1 & $-1$ &  1   & $-1$ &  1   &  $Z$  & $XY + YX$
\\
 $E$   & 2 &  0   & $-2$ &  0   &  0   & $(X, Y)$, & $(YZ, \, XZ), $ \\ 
 & & & & & & $(R_X, \, R_Y)$  & $(ZY, \, ZX)$
\\
\hline\hline
\end{tabular}
\vspace*{-3mm}
\caption{Character table of the point group $D_{2d}$. $A_1$, $A_2$, $B_1$, $B_2$, and $E$ represent irreducible representations, and ($X$, $Y$, $Z$) and ($R_X$, $R_Y$, $R_Z$) are basis functions corresponding to the polar vectors ($\bm j$, $\bm E$, and ${\bm \nabla} T$) and axial vectors (${\bm E} \times {\bm \nabla} T$ and $\bm M$), respectively.}
\label{tab:T}
\end{center}
\end{table}

Up to this point, we have examined the specific properties of various physical quantities in the material Cu$_2$WSe$_4$. From here, we show the general relations of the nonlinear Hall conductivity, expressed as a third-rank tensor, under the point group $D_{2d}$, and then discuss the properties common to the NCTE Hall conductivity $\chi_{ij}$ and the linear-response Edelstein coefficient $\alpha_{ij}$, as well as their relation to crystal chirality.

First, we consider the nonlinear component of the current induced by two generalized driving forces $\bm F$ and ${\bm F}'$ (polar vectors): $j_i = \sigma_{ijl}^{FF'} F_j F'_l $. The character table of point group $D_{2d}$, with the rotoreflection axis along the $z$-axis, is shown in Table~\ref{tab:T}. From this, it immediately follows that $\sigma_{xyz}^{FF'} = \sigma_{yxz}^{FF'}, \ \sigma_{yzx}^{FF'} = \sigma_{xzy}^{FF'}, \ \sigma_{zxy}^{FF'} = \sigma_{zyx}^{FF'}$.

Next, we consider the response to the cross product of the two driving forces: $j_i^{\rm A} = {\cal C}_{ij} ({\bm F} \times {\bm F}')_j $. This directly includes the setup of the NCTE Hall effect. From the symmetry relations with respect to $\sigma_{ijl}$ discussed above, one can immediately show that the tensor ${\cal C}_{ij}$ obeys ${\cal C}_{xx} = -{\cal C}_{yy}$ and ${\cal C}_{zz} = 0$ under point group $D_{2d}$. 

Note that the tensor $\cal C$ is connecting polar and axial vectors, $\bm j$ and ${\bm F} \times {\bm F}'$, respectively. Thus, this result can be generalized by recalling the tensor ${\cal G}$ introduced before, which includes not only the NCTE Hall effect but also the Edelstein effect as well. As stated before, the systems with ${\cal G} \neq 0_{3\times 3}$ are defined as gyrotropic systems. There are 18 such point groups, and $D_{2d}$ belongs to this class. We can further classify these gyrotropic point groups into the following three parts~\cite{GG2014,FSKI2017,HL2020,SW2025}.  

\noindent
$\bullet$ Weakly gyrotropic (3 point groups): 
\begin{align}
C_{3v} \ (3m), \ C_{4v} \ (4mm), \ C_{6v} \ (6mm).   \nonumber
\end{align}

\noindent
$\bullet$ Strongly gyrotropic, achiral (4 point groups): 
\begin{align}
C_{s} \ (m), \ C_{2v} \ (mm2), \ S_{4} \ (\bar 4), \ D_{2d} \ (\bar 4 2 m).   \nonumber
\end{align}

\noindent
$\bullet$ Strongly gyrotropic, chiral  (11 point groups): 
\begin{align}
&O \ (432), \ T \ (23), \ D_6 \ (622), \ D_4 \ (422), \ D_3 \ (32), \nonumber
\\
&D_2 \ (222), \ C_6 \ (6), \ C_4 \ (4), \ C_3 \ (3), \, C_2 \ (2), \ C_1 \ (1). \nonumber
\end{align}

\noindent
Regarding the relation to $\cal G$, we can show that the weakly gyrotropic systems do not have finite diagonal component in arbitrary coordinate: 
\begin{align}
({\rm weakly~gyrotropic}) \to {\cal G}_{ii} = 0.
\end{align}
Therefore, we get 
\begin{align}
{\cal G}_{ii} \neq 0 \to ({\rm strongly~gyrotropic}) .
\end{align} 
In case of Cu$_2$WSe$_4$, the diagonal components satisfy ${\cal G}_{xx} = -{\cal G}_{yy} \neq 0$ and ${\cal G}_{zz} = 0$ as well as ${\cal G}_{ij} = 0$ when $i\neq j$. Thus, we can classify this material to the strongly gyrotropic system. Furthermore, if the point group of a system belongs to achiral sector, it can be shown that ${\rm tr} \, {\cal G} = 0$. Namely, 
\begin{align}
({\rm achiral}) \to {\rm tr} \, {\cal G} = 0,
\end{align}
and thus
\begin{align}
{\rm tr} \, {\cal G} \neq 0 \to ({\rm chiral}).
\end{align}
Again quoting the actual case of Cu$_2$WSe$_4$, we can also confirm that ${\cal G}_{xx} = -{\cal G}_{yy}$ and ${\cal G}_{zz} = 0$, leading to ${\rm tr} \, \chi = {\rm tr} \, \alpha = 0$. In contrast, for chiral crystal structures such as Te and CoSi (which belong to $D_3$ and $T$ point groups, respectively), these traces are finite~\cite{NYY2024,NYY2025}. This means that ${\rm tr} \, {\cal G}$ can serve as a detector of chirality, justifying the name “chirality indicator.” The signs of these chirality indicators depend on the crystal chirality, and thus ${\rm tr} \, {\cal G}_{\rm L} / {\rm tr} \, {\cal G}_{\rm R} = -1$ (${\cal G}_{\rm L}$ and ${\cal G}_{\rm R}$ are gyrotropic tensors in left- and right-handed crystals, respectively) should be satisfied if we can compare in the same chemical potential. Further details are summarized in Appendix~\ref{sec:G}. 

Finally, let us look at the connection between chirality and the orbital magnetic moment. For the Edelstein magnetization and the NCTE charge and thermal Hall current, one finds 
\begin{align}
{\rm tr} \, \alpha 
&= \frac{e \tau}{\hbar} \frac{1}{V} \sum_{n\k} f(\ve_{n\k}) \, {\bm \nabla}_\k \cdot \left( {\bm m}_{n\k}^{\rm S} + {\bm m}_{n\k}^{\rm O} \right), 
\\
{\rm tr} \, \chi^{(\ell)} 
&= - \frac{e \tau}{\hbar} \frac{1}{V} \sum_{n\k} F_\ell (\ve_{n\k}) {\bm \nabla}_\k \cdot {\bm m}_{n\k}^{\rm O}. 
\label{eq:trchi}
\end{align}
Interestingly, the Berry curvature contribution in NCTE Hall effect is irrespective for the indicator. It is also notable that these indicators are expressed by the divergence of the magnetic moments. Therefore, the necessary conditions for these indicators to be finite are the existence of (i) monopole-like structures in either the spin texture or the orbital magnetic moment in the momentum space and (ii) an energy difference between the monopoles with mutually opposite monopole numbers ($\mu_5 \neq 0$). This is consistent with previous studies on the electronic structures of chiral materials~\cite{Chang2018,Tamanna2024}. In our study of Cu$_2$WSe$_4$, $\bm m_{n\k}$ (and $\bm \Omega_{n\k}$) only include the planar dipole-like structure without monopoles due to the rotoreflection symmetry. Whereas in Te and CoSi, Weyl points and multifold chiral fermions exist in momentum space, from which the divergence (singularity) of orbital magnetic moments indeed emanate~\cite{Hirayama2015,TZZ2017,Maruggi2023,Yang2023,NYY2024,NYY2025}. Moreover, the divergence of the orbital magnetic moment takes the form of an inner product between an axial and a polar vector, sharing a feature with quantities proposed in earlier works for quantifying chirality~\cite{MISH2025,OK2025}.

These properties demonstrate that the chirality and gyrotropy can be directly connected with actual transport coefficients and magnetoelectric cross-correlation responses. The chirality indicator ${\rm tr} \, {\cal G}$ clearly distinguishes gyrotropy and chirality, which can promote future experimental measurements and investigation. Moreover, a symmetry-breaking distortion applied to the achiral crystal is expected to induce an onset of the trace of the gyrotropic tensor, and the chirality indicator provides an experimental reference for detecting chiral phase transitions and for realizing engineered chirality. 

\section{Conclusions}
\label{sec:conc}

In this study, we have investigated the electronic structure and orbital properties of the achiral gyrotropic material Cu$_2$WSe$_4$, and discussed the general relationship between chirality and the orbital magnetic moment. Our analysis reveals that the orbital Edelstein effect plays a dominant role compared to the spin contribution. For both the nonlinear Hall and NCTE Hall effects, we find that the Berry curvature dipole provides the leading contribution, while the orbital magnetic moment plays a secondary role. Furthermore, we have elucidated the common features of the response coefficients characterizing the Edelstein and NCTE Hall effects, and clarified the distinctions between chiral and gyrotropic systems by introducing a “chirality indicator.” These results establish Cu$_2$WSe$_4$ as a promising candidate for novel orbitronics devices as a rectifier driven by the orbital degrees of freedom. This work also promotes the future investigation of the gyrotropic and chiral materials.

\begin{acknowledgments}
The authors thank F. Kagawa and T. Nomoto for the helpful discussions. This work is supported by JSPS KAKENHI (Grant Nos.~JP20K03835, JP21K14526, JP21K13875, JP24H00853, and JP25K07224). Parts of the numerical calculations have been done using the Supercomputer HOKUSAI BigWaterfall2 (HBW2), RIKEN. 
\end{acknowledgments}

\appendix

\section{Symmetry analysis}
\label{sec:G}

In the main text, we introduced the tensor $\cal G$ that links the axial vector and the polar vector. We described how the presence or absence of the trace of this tensor characterizes chirality, and how the presence or absence of (diagonal) components reflects the nature of gyrotropy. Here, we summarize the detailed properties of $\cal G$~\cite{GG2014,FSKI2017,HL2020,SW2025}.

\subsection{Nongyrotropic point groups}

We first list up the centrosymmetric point groups:
\begin{align}
&C_i \ (\bar 1), \ C_{2h} \ (2/m), \ D_{2h} \ (mmm), \nonumber 
\\ 
&C_{4h} \ (4/m), \ D_{4h} \ (4/mmm), \ C_{3i} \ (\bar 3), \nonumber 
\\ 
&D_{3d} \ (\bar 3/m), \ C_{6h} \ (6/m), \ D_{6h} \ (6/mmm), \nonumber
\\ 
&T_h \ (m\bar 3), \ O_h \ (m\bar 3 m) , 
\end{align}
where the second-order responses as well as Edelstein effect are prohibited.  

Noncentrosymmetric point groups allow the second-order responses. However, the point groups
\begin{align}
C_{3h} \ (\bar 6), \ D_{3h} \ (\bar 6 m2), \ T_d \ (\bar 4 3 m),   
\end{align}
do not exhibit finite gyrotropic tensor as well: ${\cal G} = 0_{3\times 3}$. Namely, Edelstein effect as well as NCTE Hall effect should be absent. 

\subsection{Gyrotropic point groups}

Other 18 point groups belong to the gyrotropic class, which allows ${\cal G} \neq 0_{3\times 3}$. They are classified as follows:

\subsubsection{Weakly gyrotropic} 
Weakly gyrotropic point groups satisfy ${\cal G}_{ii} = 0$ for arbitrary coordinate. The form of gyrotropic tensor is 
\begin{align}
{\cal G}_{C_{3v,4v,6v}} = 
\begin{pmatrix}
 0 & {\cal G}_{xy} & 0
\\
 -{\cal G}_{xy} & 0 & 0
\\
 0 & 0 & 0
\end{pmatrix}
\end{align}

\subsubsection{Strongly gyrotropic, achiral} 

${\cal G}_{ii} \neq 0$ is a sufficient condition to be strongly gyrotropic, but always satisfy ${\rm tr} \, {\cal G} = 0$ in case of the achiral point groups. The gyrotropic tensor for each point group is
\begin{align}
{\cal G}_{C_s} &= 
\begin{pmatrix}
 0 & {\cal G}_{xy} & 0
\\
 {\cal G}_{yx} & 0 & {\cal G}_{yz}
\\
 0 & {\cal G}_{zy} & 0
\end{pmatrix} ,
\\
{\cal G}_{C_{2v}} &= 
\begin{pmatrix}
 0 & {\cal G}_{xy} & 0
\\
 {\cal G}_{yx} & 0 & 0
\\
0 & 0 & 0
\end{pmatrix} ,
\\
{\cal G}_{S_4} &= 
\begin{pmatrix}
 {\cal G}_{xx} & {\cal G}_{xy} & 0
\\
 -{\cal G}_{xy} & -{\cal G}_{xx} & 0
\\
 0 & 0 & 0
\end{pmatrix} ,
\\
{\cal G}_{D_{2d}} &= 
\begin{pmatrix}
 {\cal G}_{xx} & 0 & 0
\\
 0 & -{\cal G}_{xx} & 0
\\
 0 & 0 & 0
\end{pmatrix} .
\end{align}
Note that, since ${\cal G}_{xy} \neq {\cal G}_{yx}$ for the point groups $C_s$ and $C_{2v}$, the diagonal components can be finite in general. ${\cal G}_{D_{2d}}$ corresponds to the case of Cu$_2$WSe$_4$. 
 
\subsubsection{Strongly gyrotropic, chiral} 

${\rm tr} \, {\cal G} \neq 0$ is a sufficient condition for the chiral system. The gyrotropic tensor for each point group is
\begin{align}
{\cal G}_{O,T} &= 
\begin{pmatrix}
 {\cal G}_{xx} & 0 & 0
\\
 0 & {\cal G}_{xx} & 0
\\
 0 & 0 & {\cal G}_{xx}
\end{pmatrix},
\\
{\cal G}_{D_{6,4,3}} &= 
\begin{pmatrix}
 {\cal G}_{xx} & 0 & 0
\\
 0 & {\cal G}_{xx} & 0
\\
0 & 0 & {\cal G}_{zz}
\end{pmatrix},
\\
{\cal G}_{D_2} &= 
\begin{pmatrix}
 {\cal G}_{xx} & 0 & 0
\\
 0 & {\cal G}_{yy} & 0
\\
 0 & 0 & {\cal G}_{zz}
\end{pmatrix},
\\
{\cal G}_{C_{6,4,3}} &= 
\begin{pmatrix}
 {\cal G}_{xx} & {\cal G}_{xy} & 0
\\
 -{\cal G}_{xy} & {\cal G}_{yy} & 0
\\
 0 & 0 & {\cal G}_{zz}
\end{pmatrix},
\\
{\cal G}_{C_{2}} &= 
\begin{pmatrix}
 {\cal G}_{xx} & 0 & {\cal G}_{xz}
\\
 0 & {\cal G}_{yy} & 0
\\
 {\cal G}_{zx} & 0 & {\cal G}_{zz}
\end{pmatrix},
\\
{\cal G}_{C_1} &= 
\begin{pmatrix}
 {\cal G}_{xx} & {\cal G}_{xy} & {\cal G}_{xz}
\\
 {\cal G}_{yx} & {\cal G}_{yy} & {\cal G}_{yz}
\\
 {\cal G}_{zx} & {\cal G}_{zy} & {\cal G}_{zz}
\end{pmatrix}.
\end{align}
Te and CoSi belong to the point groups $D_3$ and $T$, respectively, consistent with the fact that they possess the chiral crystal structures. 

%\bibliography{Nonlinear}

%

\end{document}